\documentclass[a4paper,12pt]{article}

\def\displayandname#1{\rlap{$\displaystyle\csname #1\endcsname$}%
                      \qquad \texttt{\char92 #1}}

\usepackage{epsfig}
\usepackage{amsfonts}

\begin{document}

\title{"Dressing" and Haag's theorem\footnote{This article was published
originally as JINR preprint P2-7210, Dubna, 1973. It is translated
from Russian (with minor additions and corrections) by Eugene
Stefanovich. The translation was approved by the author Mikhail
Shirokov. Mike Mowbray helped to proofread the English version.}}

\author{M. I. Shirokov\footnote{ Joint institute for nuclear research, Dubna}}

\maketitle

\begin{abstract}
It is demonstrated that the "dressed particle" approach to
relativistic local quantum field theories does not contradict
Haag's theorem. On the contrary, "dressing" is the way to overcome
the difficulties revealed by Haag's theorem.
\end{abstract}

\section{ Introduction} \label{sc:intro}

Two corpuscular interpretations are well-known for relativistic
local theories of interacting fields: i.e., those in terms of "bare"
and \textit{in-out} particles. The first is the corpuscular
interpretation of free fields, though used in the case when
interactions between the fields are turned on. Its drawbacks are
known. In order to formulate scattering problems correctly one uses
\textit{in-out} operators. However, the task of determining them
in a given Lagrangean theory coincides, in fact, with the task
of diagonalizing of the full Hamiltonian H. Usually, the
\textit{in-out} operators are not calculated, but postulated.

In this work, a corpuscular interpretation in terms of "dressed"
particles is discussed. A "dressed" particle is to be understood as
a particle described by creation-annihilation operators
$\alpha^{\dag}, \alpha$ with the following properties:

\begin{itemize}
\item[a)]The spectrum of indices enumerating $\alpha^{\dag},
\alpha$ should be the same as for "bare" operators $a^{\dag}, a$.
The commutation relations for $\alpha^{\dag}, \alpha$ are also
canonical:

\begin{eqnarray}
[\alpha_{\mathbf{p}}, \alpha^{\dag}_{\mathbf{p}'}] =
\delta(\mathbf{p-p}') \label{eq:1}
\end{eqnarray}

The usual Fock representation of (\ref{eq:1}) is adopted, so that
a particle number operator exists, as is necessary for
the corpuscular interpretation.
\item[b)] The no-particle vector $\Omega$ of this representation ($\alpha_{\mathbf{p}} \Omega = 0$ for all
$\mathbf{p}$) coincides with the physical vacuum of the theory (the
eigenvector of $H$ with the lowest energy).
\item[c)] The one-particle states $\alpha^{\dag}_{\mathbf{p}}
\Omega$ must be also eigenvectors of $H$.
\end{itemize}

These fundamental requirements (cf. \cite{GS}) are usually augmented
by a number of additional ones. For example, the state
$\alpha^{\dag}_{\mathbf{p}} \Omega$ should be an eigenstate of the
total momentum, and should possess definite quantum numbers, like
charge, parity, etc. (see \cite{Shirokov-72} and Sect. 3.1 in
\cite{Shirokov4}). However, for the present work only properties a),
b) are significant.

There exist other definitions of "dressing". See the end of section
\ref{section1}. Note that \textit{in-out} operators satisfy
requirements a), b), c), but in addition they also
have the property of stationarity for all states of the form
$\alpha^{\dag}_{\mathbf{p}_1} \ldots \alpha^{\dag}_{\mathbf{p}_n}
\Omega$. For more about the connection between the \textit{in-out} and
"dressed" operators see section \ref{section1} below. Section
\ref{section1} is mainly devoted to demonstrating
non-locality of the "dressed field". This property of "dressing" is
used in a very substantial way in section \ref{section2}, where the
main statement of this work is proved: that one cannot reject the
possibility of a "dressed" corpuscular interpretation of
relativistic local field theory on the basis of Haag's theorem.
On the contrary, "dressing" is the way to overcome the
difficulties revealed by Haag's theorem.

\section{"Dressing" and non-locality} \label{section1}

Here we present a simplified version of the formal "dressing"
procedure, due to Faddeev \cite{Faddeev}. This procedure
is applicable to any relativistic local field theory.

Let us assume, for definiteness, that the interaction Hamiltonian
density is the product of three field operators (quantum
electrodynamics $\overline{\psi} \gamma_{\mu} \psi A_{\mu}$, Yukawa
interaction $\overline{\psi}  \psi \phi$, etc.). The annihilation
operators of "bare" particles (electrons, photons, nucleons, mesons)
are denoted by $a_{\mathbf{p}}$. The corresponding no-particle
vector $\Omega_0$ ($a_{\mathbf{p}}\Omega_0 = 0$ for all
$\mathbf{p}$) is an eigenvector of the free part $H_0$ of the
Hamiltonian. However, it is not an eigenvector of the total
Hamiltonian $H$, because there are interaction terms that contain
products of only (three) creation operators. The interaction terms
containing products of two creation operators and one annihilation
operator (we call them terms of type (2,1)) do not allow
the one-particle states $a^{\dag}_{\mathbf{p}}\Omega_0$ to be
eigenvectors of $H$. Other interaction terms (of type (0,3) and
(1,2)) are Hermitian conjugate to those just mentioned.

Instead of $a_{\mathbf{p}}$, let us introduce new operators
$\alpha_{\mathbf{p}}$, which are related to $a_{\mathbf{p}}$ by a
formally unitary transformation

\begin{eqnarray}
a_{\mathbf{p}} = W \alpha_{\mathbf{p}}W^{\dag}, \mbox{ }
a^{\dag}_{\mathbf{p}} = W\alpha^{\dag}_{\mathbf{p}}W^{\dag}
\label{eq:2}
\end{eqnarray}

\noindent (so that properties a) from the Introduction are satisfied).
One simplest example: $W = \exp \{ \frac{1}{2} \int d^3p \chi
(|\mathbf{p}|)[\alpha_{\mathbf{p}}\alpha_{-\mathbf{p}} -
\alpha^{\dag}_{\mathbf{p}}\alpha^{\dag}_{-\mathbf{p}}] \}$
corresponds to the linear transformation

\begin{eqnarray}
a_{\mathbf{p}} = \cosh \chi (|\mathbf{p}|) \alpha_{\mathbf{p}} +
\sinh \chi (|\mathbf{p}|) \alpha^{\dag}_{-\mathbf{p}} \label{eq:3}
\end{eqnarray}

\noindent  Let $H(a^{\dag}, a)$ denote the full Hamiltonian
expressed in terms of "bare" operators $a^{\dag}, a$. If in this
expression we transform from $a^{\dag},a$ to $\alpha^{\dag},\alpha$
we obtain the full Hamiltonian as a function of $\alpha^{\dag}, \alpha$:

\begin{eqnarray}
H (a^{\dag}, a) = H(W \alpha^{\dag}W^{\dag}, W \alpha W^{\dag}) = W
H(\alpha^{\dag}, \alpha ) W^{\dag} =  K(\alpha^{\dag}, \alpha )
\label{eq:4}
\end{eqnarray}

\noindent  (where we used formulas of the type $f( W \alpha
W^{\dag}) = W f( \alpha ) W^{\dag}$). Now we need to construct  $W$
such that the full transformed Hamiltonian $K(\alpha^{\dag}, \alpha
)$ would not contain "bad" terms of type (3,0), (2,1), and generally
of types $(m,0)$ and $(m,1)$ with $m \geq 2$. It is precisely these
"bad" terms which prevent the no-particle vector $\Omega$ and the
vectors $\alpha^{\dag}_{\mathbf{p}} \Omega$ from being eigenvectors
of
 $K(\alpha^{\dag}, \alpha ).$\footnote{Note that terms of type
(1,1) are allowed. The free part of $K$ is composed of them.}

The operator $W$ is constructed in the following way. Let $H
(a^{\dag}, a) = H_0 (a^{\dag}, a) + \lambda V (a^{\dag}, a)$, where
$\lambda$ is a small coupling constant. Then we represent $W$ in the
form $\exp R (\alpha^{\dag}, \alpha )$, where the anti-Hermitian
operator $R$ has the form $R = \sum_n \lambda^n R_n$, $R_n^{\dag} =
-R_n$.\footnote{This representation simplifies the procedure
suggested in \cite{Faddeev}.}

To determine $K(\alpha^{\dag}, \alpha ) = W
H(\alpha^{\dag}, \alpha ) W^{\dag}$ we use the formula

\begin{eqnarray*}
e^{A}Be^{-A} = B + [A,B] + \frac{1}{2} [A,[A,B]] + \frac{1}{3!}
[A,[A,[A,B]]] + \ldots 
\end{eqnarray*}

\noindent  to obtain a power series in $\lambda$

\begin{eqnarray}
K(\alpha^{\dag}, \alpha ) &=&  K_0(\alpha^{\dag}, \alpha ) + \lambda
K_1(\alpha^{\dag}, \alpha ) + \lambda^2 K_2(\alpha^{\dag}, \alpha )
+ \ldots; \label{eq:5} \\
K_1 &=& [R_1, H_0 ] + V; \mbox{  }  K_2 = [R_2, H_0 ] + [R_1, V ] +
\frac{1}{2} [R_1, [R_1, H_0 ]] + \ldots \label{eq:6}
\end{eqnarray}

\noindent  $K_1$ contains interaction terms $V$. All are
"bad" in the case of the three-operator interaction. By an appropriate
choice of $R_1$ we can make $K_1$ zero. To do this, we choose
$R_1$ to be a three-operator expression of the same structure as $V$,
but with other coefficient functions. Then $[R_1, H_0 ]$ is
also a three-operator expression, which can be made equal to $-V$ by
an appropriate choice of these coefficient functions. (Note that the
masses of particles should be such that the decay of one particle
into two is impossible).

After finding $R_1$, we can calculate all terms in $K_2$ except
$[R_2, H_0]$, see (\ref{eq:6}). There are "bad" terms among them. To
find them, we perform normal ordering of terms $[R_1, V ]$
and $ [R_1, [R_1, H_0 ]] = - [R_1, V ]$. (I.e: we move all creation
operators to the left of any annihilation operators by using the
commutation relations (\ref{eq:1})). Thus we obtain "bad" terms
of types (2,0), (4,0), (3,1). If we take $R_2$ as a superposition of
terms of the same types, then the corresponding coefficient
functions in $R_2$ can be chosen such that $[R_2, H_0]$ compensates
"bad" terms from $[R_1, V ]$. Similarly, one can delete "bad" terms
from $K_n$ with any $n$ \cite{Faddeev}.

Then in $K_2$, $K_3$, etc, only "good" terms are left, e.g: of
type $\alpha^{\dag}\alpha^{\dag}\alpha\alpha$. They describe
interactions that lead to scattering and more complicated reactions.
Let us now demonstrate that these interactions are non-local.

We formally introduce the Heisenberg operator of the "dressed field"

\begin{eqnarray}
A(\mathbf{x}, t) = (2 \pi)^{-3/2} \int d^3p \frac{1}{\sqrt{2
E_{\mathbf{p}}}} [e^{-iE_{\mathbf{p}}t + i
\mathbf{px}}\alpha(\mathbf{p}, t) + e^{iE_{\mathbf{p}}t - i
\mathbf{px}}\alpha^{\dag}(\mathbf{p}, t)]\label{eq:7}
\end{eqnarray}

\noindent  which is built in the usual manner from the Heisenberg
"dressed" creation-annihilation operators $\alpha(\mathbf{p}, t) =
\exp(iHt)\alpha_{\mathbf{p}} \exp(-iHt)$. As with $\alpha^{\dag},
\alpha$, the symbol $A$ represents a set of operators for all fields
in the theory under consideration.

If the four-operator part of the interaction were local, for example
of type $gA^4(x)$, then in addition to
$\alpha^{\dag}\alpha^{\dag}\alpha\alpha$
the Hamiltonian $K$ would necessarily also contain terms of the type
$\alpha^{\dag}\alpha^{\dag}\alpha^{\dag}\alpha^{\dag}$  and
$\alpha^{\dag}\alpha^{\dag}\alpha^{\dag}\alpha$. However, such "bad"
terms were removed from $K$.

The non-locality referred to above means this: if $A(x) =
A(\mathbf{x}, t)$ satisfies an equation of the type $(\Box + m^2)
A(x) = J(x)$, then the current $J(x)$ is non-local in the sense that

\begin{eqnarray}
[J(x), A(y)] \neq 0 \mbox{ } when \mbox{ } (\mathbf{x-y})^2 > (x_0 -
y_0)^2 \label{eq:8}
\end{eqnarray}

\noindent We conclude this section with three remarks.

\begin{itemize}
\item[1.] For local interaction, Faddeev's
procedure leads to divergences. For example, normal ordering of
$[R_1, V]$ creates terms of type $\int d^3p \Delta (\mathbf{p})
\alpha^{\dag}\alpha$, where $\Delta (\mathbf{p})$ is given by a
divergent integral. These terms are corrections to the free part
of the Hamiltonian (of order $\lambda^2$), and $\Delta
(\mathbf{p})$  is a correction to the energy $E_{\mathbf{p}} =
\sqrt{\mathbf{p}^2 + m^2}$. Therefore, it is necessary to introduce
a momentum cutoff and add renormalization counterterms to the
original interaction.

However, even with these improvements, the expression $\exp
R(\alpha^{\dag}\alpha)$ is not an operator, as it fails to map
vectors of the Hilbert space in the Fock representation of
operators $\alpha$ to vectors in the same space. It is known,
for instance, that even the simplest $W$ corresponding to the
transformation (\ref{eq:3}) fails in this respect.
(See \cite{Haag} page 19 and \cite{Berezin}, \S 4).
However, expressions of this kind can be
given a mathematical meaning by adopting the algebraic point of
view presented in \cite{Weidlich} and \cite{Shirokov-72}.

\item[2.] The "dressing" procedure described above enables us to discuss
the question of the connection between the "dressed" and
\textit{in-out} operators raised in \cite{GS}.

The procedure of finding \textit{in-out} operators, which is very
similar to the Faddeev's procedure, was suggested simultaneously by
Weidlich \cite{Weidlich}. It consists of deleting \emph{all} the
interaction terms in $K_n$, not just "bad" ones, so that in terms of
\emph{in} operators the full Hamiltonian $H$ must obtain the free
form. (It is presumed that this operator does not have bound
states). This is equivalent to finding all eigenstates of $H$, see
also \cite{Friedrichs}. Meanwhile "dressing" is equivalent to the
finding only the first few eigenstates of $H$ (the vacuum vector and
one-particle states). The issue of bound states requires further
investigation (now in terms of "dressed" operators). If there are
bound states then the spectrum of indices of the "dressed" operators
differs from the spectrum of indices of the \textit{in-out}
operators. We note that Heisenberg's "dressed" operators converge
strongly, i.e., with respect to the norm,  to the \textit{in-out}
operators \cite{Haag-58, Braun-Novozhilov}.

\item[3.] Note that in the literature, the term "dressing" is often
applied to transformations which differ from the Faddeev's
$W = \exp R$ described above, so that some of the conditions a), b),
c) (see Introduction) do not hold. For example, the
exponent may include only some terms from the series $\sum_n
\lambda^n R_n$, so that $W$ can even be formally non-unitary
\cite{Glimm1, Glimm2, Hepp}.
\end{itemize}

\section {"Dressing" and Haag's theorem} \label{section2}

The procedure based on perturbation theory described above for
finding "dressed" operators was formal (since questions of
convergence were not discussed). Therefore, it is important to
consider objections against the "dressing" approach based on
Haag's theorem and other similar theorems \cite{Greenberg,
Lopuszanski2}.

Haag's theorem exists in its original form (see \S 4 in
\cite{Haag}, \S 6 in \cite{Wightman}, and \cite{Guenin}) and in the
form of Hall-Wightman, which contains a large number of assumptions
(see books \cite{PCT, Bogolyubov}).

Let us first demonstrate that "dressing" is one method for
overcoming the difficulty raised by the original Haag theorem.
To do that, we present a proof of this theorem in the framework
of the Lagrangean formalism, which is somewhat different from the
proof in \cite{Haag, Wightman, Guenin}. We start with the proof
of the following lemma.

\bigskip

\textbf{Lemma.} Suppose we have a Euclidean (i.e., translational
and rotational) invariant field theory written in terms of
creation-annihilation operators $a^{\dag}_{\mathbf{p}}, a_{\mathbf{p}}$.
Suppose also that a Fock representation of these operators in the
Hilbert space $\mathcal{H}_0$ with the no-particle vector $\Omega_0$
is given. Then, in $\mathcal{H}_0$ there is a unique normalizable
eigenstate of the total momentum operator $\mathbf{P}$, and this state
coincides with $\Omega_0$.

The proof consists in analyzing all eigenstates of $\mathbf{P}$. It
is known that for any interaction the operator $\mathbf{P}$ has the
free form $P_j = \int d^3p p_j a^{\dag}_{\mathbf{p}}
a_{\mathbf{p}}$.\footnote{ We have in mind "instant form" theories
\cite{Dirac}, where time is a parameter (the states are given at a
fixed time instant).} $\mathbf{P}$ is one of the generators of the
Euclidean subgroup, so its spectrum should be continuous. Any vector
from $\mathcal{H}_0$ can be expanded in the basis $\Omega_0,
a^{\dag}_{\mathbf{p}} \Omega_0, \ldots, a^{\dag}_{\mathbf{p}_1}
\ldots a^{\dag}_{\mathbf{p}_0}\Omega_0, \ldots $. All of them are
eigenvectors of $\mathbf{P}$, but only the vector $\Omega_0$ is
normalizable. All other vectors are non-normalizable. Their
arbitrary superposition

\begin{eqnarray}
\sum_{n=1}^{\infty} \int d^3p_1 \ldots d^3p_n F(\mathbf{p}_1,
\ldots, \mathbf{p}_n) a^{\dag}_{\mathbf{p}_1} \ldots
a^{\dag}_{\mathbf{p}_n} \Omega_0  \label{eq:9}
\end{eqnarray}

\noindent is non-normalizable as well, if it is an eigenvector of
$\mathbf{P}$. Indeed, (\ref{eq:9}) is such an eigenvector if $F$
contains a delta-function of the type $\delta (\mathbf{p}_1 +
\mathbf{p}_2 + \ldots + \mathbf{p}_n - \mathbf{P})$, but then $\int
d^3p_1 \ldots d^3p_n |F(\mathbf{p}_1, \ldots, \mathbf{p}_n)|^2$
diverges.

\bigskip

\textbf{Remark} regarding the uniqueness of the normalized
eigenvector of $\mathbf{P}$: Let us write the theory in terms of
other creation-annihilation operators
$\alpha^{\dag}_{\mathbf{p}},\alpha_{\mathbf{p}}$, such that
$\mathbf{P}$ preserves its form $P_j = \int d^3p p_j
\alpha^{\dag}_{\mathbf{p}}\alpha_{\mathbf{p}}$ \cite{Shirokov-72}.
This can be achieved, for example, when the $\alpha$ are related to the
$a$ by the transformation (\ref{eq:3}). The no-particle vector $\Omega'$
(for which $\alpha_{\mathbf{p}} \Omega' = 0$) is also a
non-normalizable eigenvector of $\mathbf{P}$. For the
transformation (\ref{eq:3}) it can be shown that $\Omega'$ does not
belong to the Hilbert space constructed cyclically from the vector
$\Omega_0$ (\cite{Haag}, page 19).

\bigskip

\textbf{Theorem.} Suppose that conditions of the Lemma are satisfied
and there is a unique normalizable eigenstate of the full
Hamiltonian $H$ with lowest energy, i.e., the vacuum vector
$\Omega$. Then $\Omega$ must coincide with $\Omega_0$.

\textbf{Proof.} Since $[H, P_j]= 0$, $\Omega$ must be a common
eigenvector of $H$ and $\mathbf{P}$. However there is only one
normalizable eigenstate of $\mathbf{P}$ in $\mathcal{H}_0$, and this
eigenstate coincides with $\Omega_0$. Thus $\Omega = \Omega_0$.

\bigskip

In fact, in all local theories $\Omega$ does not coincide with the
no-particle vector of "bare" creation-annihilation operators which
diagonalize $H_0$. This means such theories violate some
assumptions of the theorem. The usual conclusion \cite{Wightman} is
to reject the Fock representation for "bare" operators and to use
instead a some "strange" representation without a particle number
operator, see, for example \S 18.3 in \cite{Barton}. It follows
from our \textbf{Remark} to the Lemma that the theorem does not
forbid the Fock representation for operators whose no-particle
vector coincides with the vacuum $\Omega$. The "dressed" operators
are exactly of this kind.

Let us now discuss the statement of O. Greenberg \cite{Greenberg}:
the Heisenberg "dressed" operators $\alpha (\mathbf{p}, t),
\alpha^{\dag} (\mathbf{p}, t)$, which describe a relativistic field
theory and realize the Fock representation of the equal-time
commutation relations $[\alpha (\mathbf{p}, t), \alpha^{\dag}
(\mathbf{p}', t)] = \delta (\mathbf{p-p'})$ must obey the free
equation of motion. Greenberg based his derivation on Haag's
theorem in Hall-Wightman form. Following \cite{PCT, Bogolyubov}
we formulate the theorem for our purposes in the following manner.

Suppose we have two field theories. One is a free theory
described by a set of free fields $A_0(x)$ acting in the Hilbert
space $\mathcal{H}_0$. The other is described by an irreducible set
of fields $A(x)$. Further, let us assume that the following
conditions are satisfied:

\begin{itemize}
\item [1)] $A(x)$ is an operator in $\mathcal{H}$ which carries a
unitary representation of translations and rotations

\begin{eqnarray}
U(\mathbf{a}, R) A(x) U^{\dag}(\mathbf{a}, R) = A(Rx + \mathbf{a})
\end{eqnarray}

and

\item[1')] Lorentz transformations

\begin{eqnarray}
U(\Lambda) A(x) U^{\dag}(\Lambda) = A(\Lambda x)
\end{eqnarray}

\noindent (these relationships are written for the particular case
of a scalar field).

\item[2)] There is a unique invariant state $U \Omega = \Omega$ in
$\mathcal{H}$.

\item[3)] There exists a unitary operator $V$, from $\mathcal{H}_0$ to
$\mathcal{H}$, such that at a time instant $t$ we have

\begin{eqnarray}
 A(\mathbf{x}, t)  = V(t)A_0(\mathbf{x}, t) V^{\dag}(t)
\end{eqnarray}

\item[4)] The spectrum of energies is bounded from below.
\end{itemize}

\noindent Then $A(x)$ is a free field.

\bigskip

As a field operator describing the interacting theory we take
the "dressed" field operator \cite{Friedrichs}. Greenberg noticed
that the unitary equivalence 3) of the fields $A(x)$ and $A_0(x)$
results from the requirement a) to the "dressed" operators (see
Introduction). Indeed, let us expand $A_0(x)$ in the usual manner using
the operators $a_0(\mathbf{p}), a^{\dag}_0(\mathbf{p})$, cf.
(\ref{eq:7}). Let $a_0, a^{\dag}_0$ also realize a Fock
representation of the canonical commutation relations with the
no-particle vector $\Omega_0 \in \mathcal{H}_0$. (Note that the
choice of one or another representation for the auxiliary operator
$A_0(x)$ is under our control). Then, according to a known theorem,
the operators $\alpha(\mathbf{p},t), \alpha^{\dag}(\mathbf{p},t)$ must
be connected to $a_0(\mathbf{p}), a^{\dag}_0(\mathbf{p})$ by a
unitary transformation\footnote{A transformation is called unitary
if it preserves the norm and maps $\mathcal{H}_0$ to the entire
space $\mathcal{H}$.} $\alpha(\mathbf{p},t) = V(t)
a^{\dag}_0(\mathbf{p}) V^{\dag}(t)$, see \cite{Gording} and \S 1.6
in \cite{Berezin}. Therefore, the same transformation connects
$A(x)$ and $A_0(x)$.

Assuming that the other conditions of Haag's theorem are satisfied,
Greenberg concluded that the "dressed field" $A(x)$ must be free.
We show that in this situation condition 1') of the theorem is
not satisfied, and therefore such a conclusion is wrong.

We first demonstrate that the local commutation relation

\begin{eqnarray}
 [A(\mathbf{x}, x_0), A(\mathbf{y}, y_0) ]  = 0 \label{eq:10}
\end{eqnarray}

\noindent  is invalid for \emph{all} $(\mathbf{x}, x_0)$ separated
by a space-like interval from $(\mathbf{y}, y_0)$. Indeed, if
(\ref{eq:10}) were true, then by applying the operator $(\nabla_x^2 +
m^2)$ to (\ref{eq:10}) we would obtain the relationship
$[J(\mathbf{x}, x_0), A(\mathbf{y}, y_0)] = 0$ when $(\mathbf{x},
x_0) \approx (\mathbf{y}, y_0)$. This contradicts the established
non-locality of the current $J$ corresponding to the operator $A$,
see (\ref{eq:8}). For another proof of the impossibility of
(\ref{eq:10}) see  \cite{Braun-Novozhilov}.

Already, the non-locality of $A(\mathbf{x}, t)$ means that Haag's
theorem in this situation cannot be proved using the
Jost-Schroer theorem (theorem 4-15 in \cite{PCT}). In
Greenberg's proof, the locality condition for the field is not used
(we have not included this condition in the assumptions of
Haag's theorem) \cite{Greenberg}.

We now demonstrate that assuming the validity of
1') for $A(\mathbf{x}, t)$ leads to a contradiction. Note that the
non-local field $A(\mathbf{x}, t)$ is such that (\ref{eq:10}) is
valid for \emph{some} $(\mathbf{x}, x_0) \approx (\mathbf{y}, y_0)$.
In particular, when $x_0 = y_0 = t$ we have

\begin{eqnarray}
 [A(\mathbf{x}, t), A(\mathbf{y}, t) ]  = 0 \label{eq:11}
\end{eqnarray}

This follows from the equal-time commutation relations
$[\alpha(\mathbf{p},t), \alpha^{\dag}(\mathbf{p'},t)] = \delta
(\mathbf{p} - \mathbf{p}')$ for the Heisenberg "dressed" operators
$\alpha(\mathbf{p},t) = \exp (iHt) \alpha_{\mathbf{p}} \exp (-iHt)$
and from the expansion (\ref{eq:7}). If the condition 1') were true,
then (\ref{eq:11}) implies the validity of
(\ref{eq:10}) for all $(\mathbf{x}, x_0) \approx (\mathbf{y}, y_0)$,
which is impossible.

In fact, from theorems proved in \cite{Lopuszanski2, Gachok}, an
analogous conclusion follows about the "dressed field": only a
non-local "dressed" field can be non-free.

I express my gratitude to V. Garchinsky for discussions of Haag's theorem.

\emph{The manuscript was received by the publishing department on
May 30, 1973.}


\begin{thebibliography}{10}

\bibitem{GS}
O.~W. Greenberg,  S.~S. Schweber.
\newblock {\em Nuovo Cim.}, \textbf{8}, 378 (1958).

\bibitem{Shirokov-72}
M.~I. Shirokov,
\newblock Preprint JINR, P2-6454, Dubna, 1972.

\bibitem {Shirokov4}
A.~V. Shebeko, M.~I. Shirokov.
 \newblock {\em Fiz. Elem. Chast. Atom.
Yadra}, \textbf{32}, 31 (2001). [English translation in \emph {Phys.
Part. Nucl.} \textbf{32}, 15 (2001)],
http://www.arxiv.org/nucl-th/0102037

\bibitem{Faddeev}
L.~D. Faddeev.
\newblock {\em Dokl. Akad. Nauk SSSR}, \textbf{152}, 573 (1963).

\bibitem{Haag}
R.~Haag.
\newblock {\em Dan. Mat. Fys. Medd.}, \textbf{29}, No. 12 (1955).


\bibitem{Berezin}
F.~Berezin.
\newblock {\em Method of second quantization}.
\newblock Nauka, Moscow, 1965.

\bibitem{Weidlich}
W.~Weidlich.
\newblock {\em Nuovo Cimento}, \textbf{30}, 809 (1963).

\bibitem{Friedrichs}
K.~Friedrichs.
\newblock {\em Perturbation of spectra of operators in Hilbert space}.
\newblock Mir, Moscow, 1969.

\bibitem{Haag-58}
R.~Haag.
\newblock {\em Phys. Rev.}, \textbf{112}, 669   (1958).

\bibitem{Braun-Novozhilov}
M.~Braun,  Yu. Novozhilov.
\newblock {\em JETP}, \textbf{39}, 1317 (1960).
\bibitem{Glimm1}
J.~Glimm.
\newblock {\em Commun. Math. Phys.}, \textbf{5}, 343 (1967).

\bibitem{Glimm2}
J.~Glimm.
\newblock {\em Commun. Math. Phys.}, \textbf{10}, 1 (1968).

\bibitem{Hepp}
K.~Hepp.
\newblock {\em Theorie de la renormalization}.
\newblock in Lectures Notes in Physics, vol. 2,. Springer-Verlag, Berlin, 1969.

\bibitem{Greenberg}
O.~Greenberg.
\newblock {\em Phys. Rev.}, \textbf{115}, 706 (1959).

\bibitem{Lopuszanski2}
J.~{\L}opusza{\`n}ski.
\newblock {\em J. Math. Phys.}, \textbf{2}, 743 (1961).

\bibitem{Wightman}
A.~Wightman.
\newblock {\em Problems in the relativistic dynamics of quantized fields}.
\newblock Nauka, Moscow, 1968.

\bibitem{Guenin}
M.~Guenin.
\newblock in Lectures in Theor. Physics, vol. 9A. Ed. Barut, Gordon and Breach,
  New York, 1967, p. 234.

\bibitem{PCT}
P.~Streater,  A.~Wightman.
\newblock {\em PCT, spin, statistics, and all that}.
\newblock Nauka, Moscow, 1968.

\bibitem{Bogolyubov}
N.~N. Bogolyubov, A.~A. Logunov, I.~T. Todorov.
\newblock {\em Fundamentals of the axiomatic approach in quantum field theory}.
\newblock Nauka, Moscow, 1969.

\bibitem{Dirac}
P.~A.~M. Dirac.
\newblock {\em Rev. Mod. Phys.}, \textbf{21}, 392 (1949).

\bibitem{Barton}
G.~Barton.
\newblock {\em Dispersion methods in field theory}.
\newblock Atomizdat, Moscow, 1968.

\bibitem{Gording}
L.~G{\"o}rding,  A.~Wightman.
\newblock {\em Proc. Nat. Acad. Sci. US}, \textbf{40}, 40 (1954).

\bibitem{Gachok}
V.~Gachok.
\newblock {\em Ukrainian Mathematical Journal}, \textbf{13}, 22 (1961).

\end{thebibliography}
\end{document}